\documentclass{PoS}

\usepackage{hyperref}
\usepackage{enumitem}
\usepackage{latexsym} % Gets \Box etc
\usepackage{amssymb}  % \gtrsim, \geqslant, etc etc: 
\usepackage{amsfonts}
\usepackage{amsmath} 
\usepackage{graphicx}
\usepackage{cite}
\usepackage{xspace}

\newcommand{\epem}{\mathrm{e^+e^-}}
\newcommand{\alphas}{\alpha_{\rm s}}
\newcommand{\jpsi}{{\rm J}/\psi}
\newcommand{\pt}{p_{\rm T}}
\newcommand{\alphasmZ}{\alphas(m_{_{\rm Z}})}
\newcommand{\lqcd}{\Lambda_{_{\rm QCD}}}

\newcommand{\sqrts}{\sqrt{\rm s}}

\providecommand{\bbbar}{b\overline{b}}
\providecommand{\ttbar}{t\overline{t}}

\newcommand{\pythia}{\textsc{pythia}}
\newcommand{\sherpa}{\textsc{sherpa}}
\newcommand{\madgraph}{\textsc{MadGraph}}
\newcommand{\herwig}{\textsc{herwig}}
\newcommand{\powheg}{\textsc{powheg}}
\newcommand{\resbos}{\textsc{resbos}}

\newcommand*{\eg}{e.g.}

\title{Experimental QCD summary (ICHEP 2020)}
\ShortTitle{Experimental QCD summary}

\author{David~d'Enterria\\
CERN, EP Department, CH-1211 Geneva 23, Switzerland\\
E-mail: David.d'Enterria@cern.ch}

\abstract{This writeup summarizes the main experimental studies of the strong interaction, theoretically described by quantum chromodynamics (QCD), that were presented during the ICHEP-2020 conference. The latest results, measured mostly in p-p collisions at the LHC, are categorized in seven broad topics:
\begin{enumerate}[label=(\roman*)]
    \item Extractions of the strong coupling constant $\alphasmZ$;
    \item Comparison of data to fixed-order (N$^{\rm n}$LO) perturbative QCD calculations;
    \item Determinations of parton distribution functions (PDFs); 
    \item Comparison of data to resummed (N$^{\rm n}$LL) pQCD calculations; 
    \item Parton showering and jet substructure analyses; 
    \item Semihard (double parton interactions, multiparton interactions, hard diffraction), and soft (elastic and diffractive) QCD scatterings; %and 
    \item Studies of parton hadronization in $\epem$ and p-p collisions.
\end{enumerate}
}

\FullConference{%
  40th International Conference on High Energy physics - ICHEP2020\\
  July 28 - August 6, 2020\\
  Prague, Czech Republic (virtual meeting)
}

\begin{document}

\maketitle

%%%%%%%%%%%%%%%%%%%%%%%%%%%%%%%%%%%%%%%%%%%%%%%%%%%%%%%%%%%%%%%%%%%%%%%%%%%%%%%%%%%%%%%%%%%%%%%%%%%%%%%%%%%%
\section{Introduction}

Although the study of quantum chromodynamics (QCD) is not {\it per se} the main driving force behind current and future high-energy colliders, an accurate and precise understanding of QCD is fundamental to extract physics signals, and to control their corresponding backgrounds, in proton-proton collisions at the energy frontier. %, as well as for many high-precision studies at future $\epem$ machines. 
The strong interaction directly impacts most of the experimental and theoretical studies of the Standard Model (SM) and beyond (BSM) at the CERN Large Hadron Collider (LHC) as well as at future colliders. The important role of QCD can be recognized in a multitude of physics topics and observables:
\begin{itemize}
\item An accurate description of the production cross sections of large-$\pt$ and heavy particles, as well as of their hadronic decays, through fixed-order N$^{\rm n}$LO and resummed N$^{\rm n}$LL calculations, is fundamental for any precision SM study or BSM searches at the LHC~\cite{Amoroso:2020lgh}.
\item High-precision parton distribution functions (PDFs) are required at the LHC for searches for BSM phenomena at high parton fractional momenta $x$, for precision W, Z, and H bosons studies at mid $x$, and for beyond-DGLAP studies in the low-$x$ parton saturation regime~\cite{Gao:2017yyd}.
\item The QCD coupling $\alphas$ directly affects the calculation of all particle production cross sections and hadronic decays at the LHC, chiefly contributing to parametric uncertainties of theoretical computations of many Higgs-related, %(gluon-gluon\,$\to$\,H, and associated H+$\ttbar$ cross sections, H\,$\to\ccbar,gg$ decays,...), 
as well as electroweak and top-quark, quantities~\cite{dEnterria:2019its}.
%(W and Z bosons cross sections and decays, $\ttbar$ cross sections,...).
\item A good understanding of the soft and collinear dynamics of parton showers~\cite{Dreyer:2018nbf}, as probed \eg\ through jet substructure studies~\cite{Larkoski:2017jix}, is basic for all precision-SM measurements and BSM searches involving jets. Examples include heavy-, light-quark, and gluon separation, boosted (di)jet topologies, and dijet resonance decays.
\item An accurate description of semihard QCD (multiple hard parton interactions, underlying event,...) that dominate the bulk of the cross section at the LHC, is fundamental to improve the Monte Carlo (MC) p-p event generators through parametric tuning~\cite{Khachatryan:2015pea}.
\item Non-perturbative QCD phenomena (such as colour reconnection, intrinsic parton $k_{\rm T}$, hadronization, beam remnants,...) percolate through all hadronic final-states, impacting in particular \eg\ the extraction of key SM parameters such as the W~\cite{Argyropoulos:2014zoa} and top quark~\cite{CarloniCalame:2016ouw} masses.
\end{itemize}
This writeup summarizes the main experimental QCD results presented at the ICHEP-2020 conference, organized in seven categories broadly corresponding to the topics listed above. The physics of many-body QCD, studied via high-energy heavy-ion collisions, is summarized in Ref.~\cite{HIexp}.
%Multiple recent measurements exists that have improved our understanding of the quark and gluon dynamics. This summary contribution categorizes them along the following generic topics:

%%%%%%%%%%%%%%%%%%%%%%%%%%%%%%%%%%%%%%%%%%%%%%%%%%%%%%%%%%%%%%%%%%%%%%%%%%%%%%%%%%%%%%%%%%%%%%%%%%%%%%%%%%%%
%\section{Extractions of $\alphasmZ$}
\section{Extractions of the QCD coupling}

\hspace{-0.5cm}The QCD coupling strength at the reference scale of the Z mass $\alphasmZ$ is one of the fundamental SM parameters. Its value not only chiefly impacts the theoretical calculations of all scattering and decay processes involving real and/or virtual quarks and gluons, but it also plays a role \eg\ in the stability of the electroweak vacuum~\cite{Alekhin:2012py}. The current world average of $\alphasmZ = 0.1179 \pm 0.0010$ is known to within a 0.8\% precision~\cite{Tanabashi:2018oca}, the worst (by orders of magnitude) 
\begin{figure}[htbp!]
\centering
\raisebox{8pt}{\includegraphics[width=0.55\columnwidth,height=5.3cm]{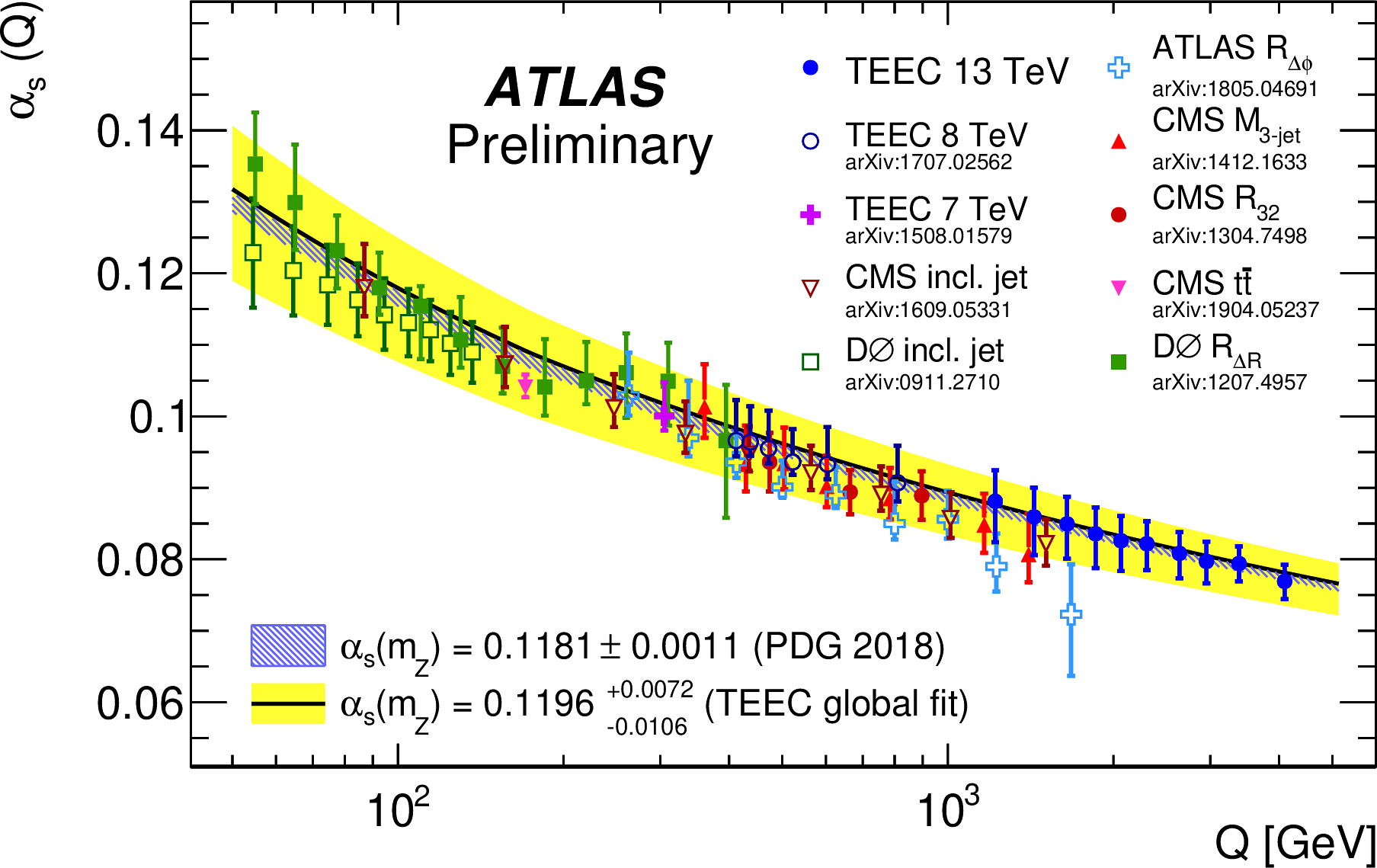}}
\hspace{0.1cm}
\includegraphics[width=0.43\columnwidth]{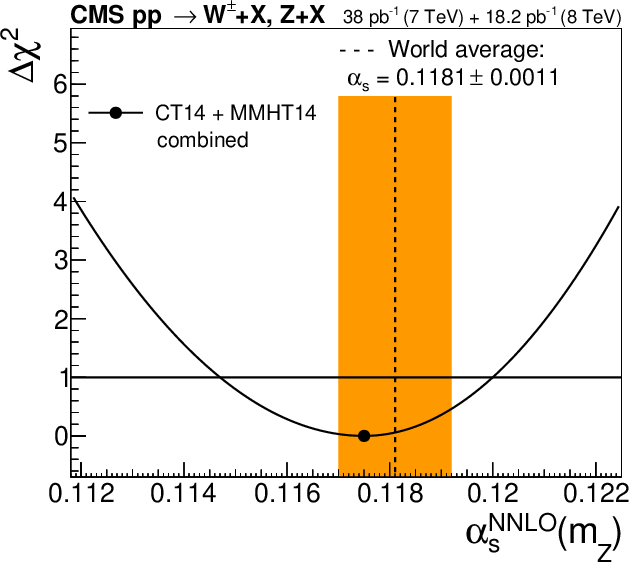}
\caption{Left: Values of $\alphas(Q)$ extracted by ATLAS from TEEC measurements in the range $Q\approx 1$--4\,TeV compared to lower $Q$ results~\cite{Loch/ATLAS}. Right: Value of $\alphasmZ$ extracted by CMS from 12 measurements of inclusive W$^\pm$ and Z cross sections (parabola) compared to the current world average (orange band)~\cite{Gorbunov/CMS,Sirunyan:2019crt}.}
\label{fig:alphas}
\end{figure}
of any other gauge coupling. Such a lack of precision propagates into dominant parametric uncertainties for many theoretical calculations, and new independent $\alphasmZ$ extractions are needed in order to further refine its value~\cite{dEnterria:2019its}. The following new QCD coupling measurements were shown at the conference:
\begin{itemize}
\item ATLAS extracted $\alphasmZ = 0.1196\pm 0.0004\,({\rm exp})\,^{+0.0072}_{-0.0106}$\,(theo) at NLO accuracy, via measurements of transverse energy-energy correlations (TEEC) and their associated azimuthal asymmetries %(ATEEC) 
in multijet events~\cite{Loch/ATLAS}. Despite its large (8\%) theoretical uncertainty, this study has significantly probed the running of $\alphas(Q)$ up to 4 TeV for the first time (Fig.~\ref{fig:alphas}, left).
\item CMS measured $\alphasmZ =0.1175 \pm0.0026$ from a global comparison of 12 precise measurements of W$^\pm$ and Z inclusive cross sections to NNLO predictions (Fig.~\ref{fig:alphas}, right)~\cite{Gorbunov/CMS,Sirunyan:2019crt}. Such measurements have been further combined with 22 other similar ATLAS and LHCb results to derive a value of $\alphasmZ = 0.1188^{+0.0019}_{-0.0013}$ with a competitive 1.3\% uncertainty~\cite{dEnterria:2019aat}.
\item Using top-quark data, a value of $\alphasmZ =0.1135^{+0.0021}_{-0.0017}$ at NLO accuracy has been derived by CMS via a combined fit of multiple $\ttbar\,+\,$N-jet differential cross sections~\cite{Bakas/CMS,Sirunyan:2019zvx}.
\item The H1 and ZEUS experiments have derived $\alphasmZ =0.1150 \pm0.0029$ at NNLO accuracy via a global fit of jets measurements in e-p collisions~\cite{Sarkar/HERA}. The inclusion of jet data leads to a larger (and more precise, $\pm$2.5\% uncertainty) value of the QCD coupling, compared to the HERAPDF result, $\alphasmZ =0.108\pm0.010$, derived using inclusive DIS data alone.
\end{itemize}
All these studies will hopefully help to improve the $\alphasmZ$ precision in upcoming world-average determinations of this key parameter. Nonetheless, extractions with ten times smaller uncertainties, in the per-mil range, require future machines such as the LHeC~\cite{AbelleiraFernandez:2012cc} as discussed by Ref.~\cite{Gwenlan/LHeC}.

%\clearpage
%%%%%%%%%%%%%%%%%%%%%%%%%%%%%%%%%%%%%%%%%%%%%%%%%%%%%%%%%%%%%%%%%%%%%%%%%%%%%%%%%%%%%%%%%%%%%%%%%%%%%%%%%%%%
\section{Data versus fixed-order (N$^{\rm n}$LO) pQCD}

Multiple differential measurements of the production of particles at high $\pt$ and/or with large mass were \begin{figure}[htbp!]
\centering
\includegraphics[width=0.29\columnwidth]{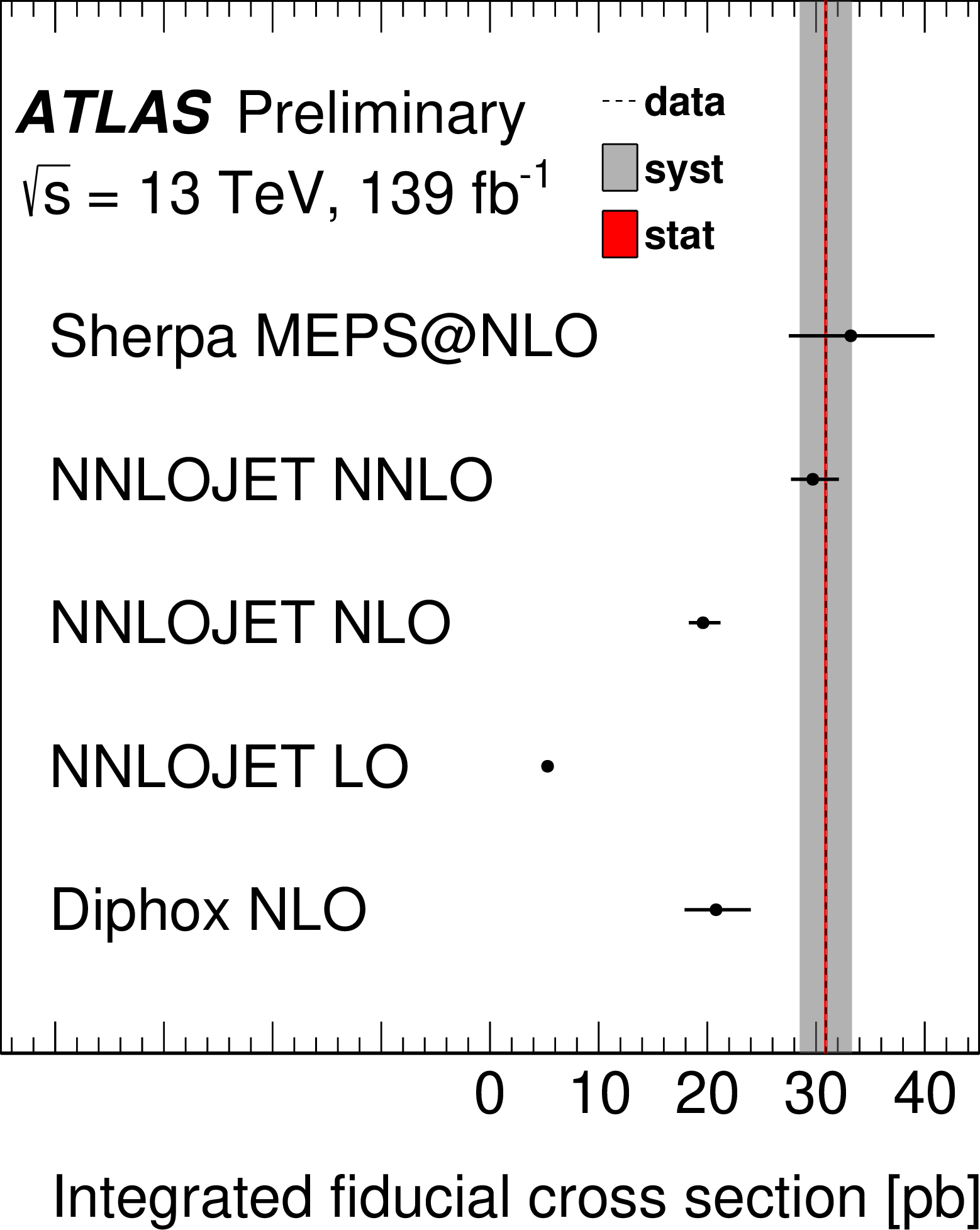}
\hspace{0.1cm}
\includegraphics[width=0.69\columnwidth]{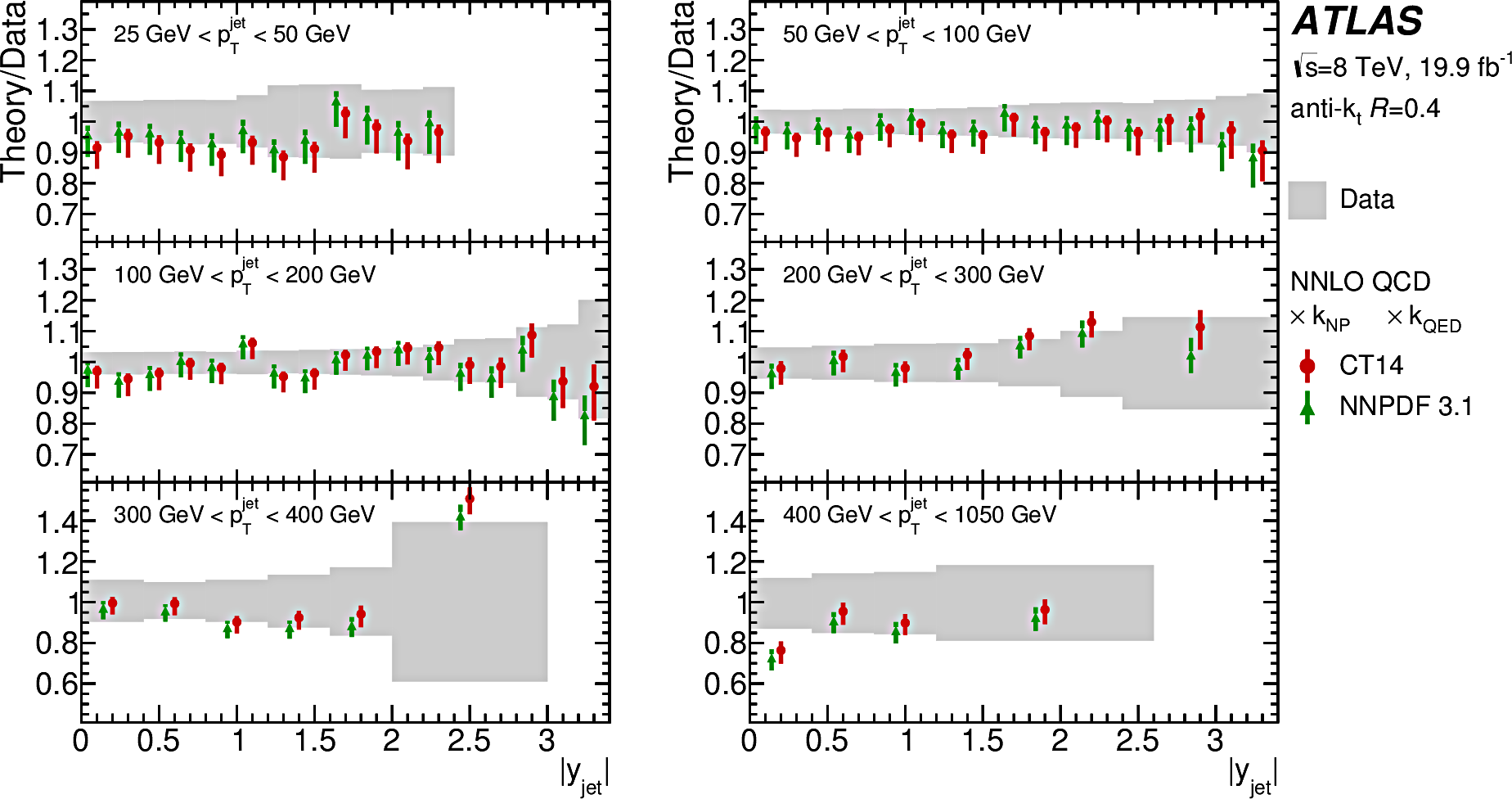}
\caption{Left: Fiducial $\gamma\gamma$ cross section measured in ATLAS compared to (N)NLO calculations~\cite{Siegert/ATLAS}. Right: Ratio of measured Z\,+\,jets cross sections over NNLO predictions, as a function of jet rapidity $|y|$ and $\pt$~\cite{Vittori/ATLAS,Aad:2019hga}.}
\label{fig:NNLO}
\end{figure}
presented during the conference: inclusive jets, heavy-quark (HQ) jets, W and Z bosons, photons,..., often also combined in pairs (W,\,Z,\,$\gamma\,+$\,jets, diphotons,...). %Z\,$+$\,HQ-jets,...). 
In the pQCD front, the current state-of-the-art are cross section calculations at next-to-next-to-leading-order (NNLO) accuracy or, for multiple hard-particle final states, NLO matrix elements (ME) combined with multileg MC parton showers (PS) such as \sherpa~\cite{Bothmann:2019yzt} or \madgraph5 (MG5)~\cite{Alwall:2011uj} plus \pythia8 (PY8)~\cite{Sjostrand:2007gs}. 
Measurements exploited to provide new PDF constraints are discussed in more detail in the next Section~\ref{sec:PDFs}. Apart from the latter results, many data were confronted to pQCD predictions:
\begin{itemize}
%\item W$^\pm$, Z, and W,\,Z\,$+$\,jets: The high-precision cross sections measured by both ATLAS~\cite{Lohwasser/ATLAS,Sutton/ATLAS} and CMS~\cite{Salvatico/CMS,Stepennov/CMS} have been exploited for new PDF constraints as discussed in the next Sec.~\ref{sec:PDFs}.
\item Jets: The latest NNLO calculations with reduced scale uncertainties~\cite{Currie:2016bfm} lead to improved data-theory accord for the $\pt$-differential ATLAS cross sections, although final states with high jet multiplicities ($N_j >3$) remain still a challenge for multileg approaches~\cite{Loch/ATLAS}.
\item Diphotons: The multiple single- and double-$\gamma$ kinematic distributions available in the $\gamma\gamma$ final state provide an excellent test-bed for NNLO and NLO+multileg %(\eg\ SHERPA) 
calculations. Whereas the total $\gamma\gamma$ rate (Fig.~\ref{fig:NNLO}, left) and the pair production at large $\phi$ are well reproduced by NNLO theory, \sherpa\ describes better the soft $\pt$ and small $\phi$ regions~\cite{Siegert/ATLAS}. 
\item Z\,$+$\,jets: The latest ATLAS Z-plus-jets spectra are well reproduced by %NLO ME+PS and 
NNLO calculations, with the NNPDF3.1 and  CT14 PDFs featuring 2--5\% differences, of the same size as the theory scale uncertainties (Fig.~\ref{fig:NNLO}, right)~\cite{Vittori/ATLAS,Aad:2019hga}. %Measurements exploitable to provide extra input for global PDF fits. . ~\cite{Malik/CMS}. 
\item Z\,$+$\,HQ jets: The production of gauge bosons with heavy quarks is an ideal channel to test multileg pQCD calculations. %4-,5-flavour (4FNS, 5FNS) multileg theory~\cite{Taheri}
The CMS Z\,$+$\,c-,\,b-jets results are well reproduced by MG5+PY8\,(2j), whereas MG5+PY8\,(4j) and \sherpa\,(4j) overestimate them~\cite{Stepennov/CMS}. %The ratios of Z\,$+$\,b-jets spectra R(b/j) and R(c/j) tend to be overestimated by the calculations
The ATLAS Z\,$+$\,1,2 b-jets cross sections (total and differential) are found to be well described by NLO-5FNS \sherpa\ and MG5 (whereas the LO 4FNS MCs largely underestimate them), but 
%Multiple differential Z+1,2 b-jets distribs. to improve higher-order  and  multileg theory: NLO-5FNS. SHERPA best. 5FNS LO MG5+PY8 better than NLO (more legs in ME+PS). But 
all models underestimate the distributions at large $m_{\bbbar}$~\cite{Vittori/ATLAS}.  %(Fig.~\ref{fig:NNLO}, right)
\end{itemize}
In summary, although %better data-theory agreement with respect to the previous generation of calculations (with one order of lower accuracy) is found for all observables, 
good data-theory agreement is found for all measurements, regions of phase space remain (\eg\ at large jet multiplicities, at large or small $\phi$ correlations for $\gamma\gamma$, or at large $m_{\bbbar}$ for Z\,$+$\,HQ jets processes) where work to incorporate higher orders of pQCD accuracy %(real legs, and virtual)
is needed. 

%\clearpage
%%%%%%%%%%%%%%%%%%%%%%%%%%%%%%%%%%%%%%%%%%%%%%%%%%%%%%%%%%%%%%%%%%%%%%%%%%%%%%%%%%%%%%%%%%%%%%%%%%%%%%%%%%%%
\section{Parton distribution functions}
\label{sec:PDFs}

\begin{figure}[htbp!]
\centering
\includegraphics[width=0.53\columnwidth,height=5.3cm]{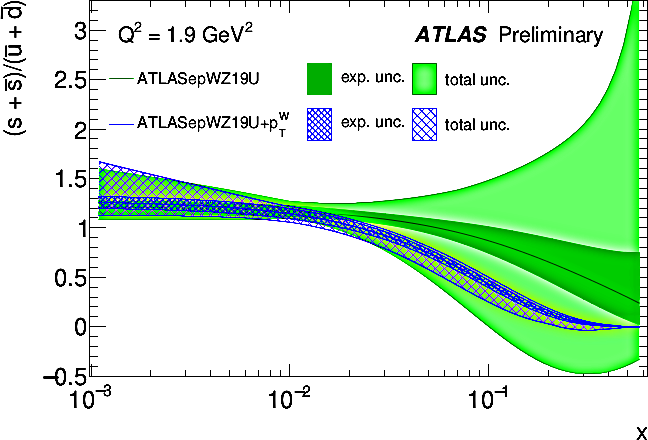}
\hspace{0.1cm}
\includegraphics[width=0.45\columnwidth,height=5.3cm]{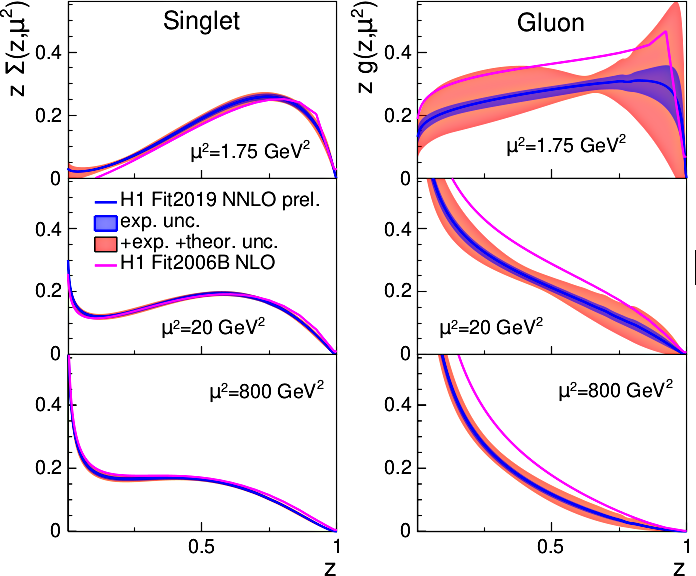}
\caption{Left: Ratio of strange over $(\bar{u}+\bar{d})$ PDFs at a scale $Q^2 = 1.9$\,GeV$^2$ determined fitting ATLAS W and Z data with and without constraints from the W\,$+$\,jets $\pt$ spectra~\cite{Sutton/ATLAS}. Right: NNLO singlet (left) and gluon (right) diffractive PDFs as a function of $z$ for three different values of the energy scale~\cite{Sarkar/HERA}.}
\label{fig:PDFs}
\end{figure}

An accurate knowledge of all parton flavour densities over the broadest $(x,Q^2)$ ranges is essential for the LHC physics program. The combination of high-precision measurements of electroweak gauge bosons, with or without jets, with NNLO pQCD predictions with percent theoretical uncertainties provides new constraints on the proton PDFs. Beyond the LHC studies, PDF results from deep-inelastic scattering (DIS) at current and future experiments were also reported during the conference. The main ICHEP'20 developments on PDFs are listed below: 
\begin{itemize}
\item Inclusive W, Z: ATLAS presented electroweak cross sections at $\sqrts = 2.76$, 13\,TeV compared to NNLO predictions with 6 different PDFs~\cite{Lohwasser/ATLAS,Aad:2019bdc}. Ratios of W$^+$, W$^-$, and Z cross sections benefit from cancellation of systematic uncertainties and provide valuable PDF constraints. The W/Z ratios give insights on the strange sea quark density, whereas the W$^+$/W$^-$ results reduce the valence $u/d$ PDF uncertainties. The HERAPDF2.0 fit tends to predict higher W/Z and W$^+$/W$^-$  ratios than measured in the data. Double differential $\pt$, $\eta$, and helicity W cross sections measured by CMS lead to 10--30\% post-fit constraints on some of the 60 Hessian NNPDF3.0 parton density sets~\cite{Salvatico/CMS,Sirunyan:2020oum}.
\item W\,$+$\,jets, W\,$+$\,charm: The W-plus-jets $\pt$ spectra incorporated into a dedicated ATLAS NNLO PDF study lead to a depleted $s$-quark density at large $x$ compared to previous fits (Fig.~\ref{fig:PDFs} left), softer (harder) sea $\bar{d}$ density at low (high) $x$, and harder (softer) valence $d$-quarks at small (high) $x$~\cite{Sutton/ATLAS}.
%\item W\,$+$\,charm: 
The comparatively depleted strange-quark PDF at large $x$ is also confirmed by the latest W\,$+$\,charm data measured by CMS~\cite{Stepennov/CMS,Sirunyan:2018hde}.% is also well reproduced by NLO predictions with depleted strange-quark PDF and strangeness suppression factor at large $x$ %agree with $\nu$-scatt. exps., disagree with previous ATLASepWZ16nnlo (central values).
 \item Photons, $\gamma+$\,jets: Multiple ATLAS and CMS differential distributions have been compared to NLO predictions~\cite{Siegert/ATLAS,Aad:2019eqv,Malik/CMS,Sirunyan:2019uya}. Since the LHC photon data have 10--15\% (both experimental and theoretical) energy-scale uncertainties, those results have led so far only to a mild impact on PDFs~\cite{dEnterria:2012kvo,Carminati:2012mm}, but stronger constraints on global PDF fits should be possible in the near future exploiting the recently available NNLO calculations~\cite{Campbell:2018wfu,Chen:2019zmr}.
\end{itemize}
\begin{itemize}
%\item $\gamma$ +jet: CMS photon+jet spectra compared to NLO pQCD with varying PDFs:~\cite{Malik/CMS} Mild differences between PDFs within uncertainties. Better exploitation to provide extra input for global PDF fits with NNLO calculations.
\item Forward HQ jets: LHCb has measured charm and bottom jet cross sections with an improved 2D BDT fit to separate flavour composition~\cite{Sestini/LHCb}. Such first-ever forward HQ jets measurements can provide novel useful constraints on the low- and  high-$x$ gluon PDF, via their incorporation into global fits with recent NNLO calculations for their cross sections~\cite{Catani:2020kkl}.
\item Diffractive PDFs: A new H1 NNLO fit of inclusive diffractive DIS data shows a 25\%~depletion of the gluon content of the pomeron compared to NLO analyses (Fig.~\ref{fig:PDFs} right)~\cite{Sarkar/HERA}. These~new dPDFs lead to a better theoretical reproduction of the HERA diffractive~dijet~cross~sections.
%\item Future DIS: Prospects for PDF studies at various future DIS facilities, including generalized PDFs at COMPASS via deeply virtual Compton scattering~\cite{COMPASS}, nuclear and polarized PDFs with the electron-ion collider (EIC)~\cite{Joosten/EIC}, and the ultimate PDF reach over many orders-of-magnitude in the $(x,Q^2)$ plane reachable at the LHeC~\cite{Gwenlan/LHeC,AbelleiraFernandez:2012cc} were presented.
\end{itemize}
Beyond the results above, prospects for parton densities studies at various future DIS facilities, including generalized PDFs at COMPASS via deeply virtual Compton scattering~\cite{COMPASS}, nuclear and polarized PDFs at the electron-ion collider (EIC)~\cite{Joosten/EIC,Accardi:2012qut}, and the ultimate PDF reach over many orders-of-magnitude in the $(x,Q^2)$ plane accessible at the LHeC~\cite{Gwenlan/LHeC,AbelleiraFernandez:2012cc}, were also reported.

%%%%%%%%%%%%%%%%%%%%%%%%%%%%%%%%%%%%%%%%%%%%%%%%%%%%%%%%%%%%%%%%%%%%%%%%%%%%%%%%%%%%%%%%%%%%%%%%%%%%%%%%%%%%
\section{Data versus resummed (N$^{\rm n}$LL) pQCD}

\begin{figure}[htbp!]
\centering
\raisebox{12pt}{\includegraphics[width=0.47\columnwidth,height=6.4cm]{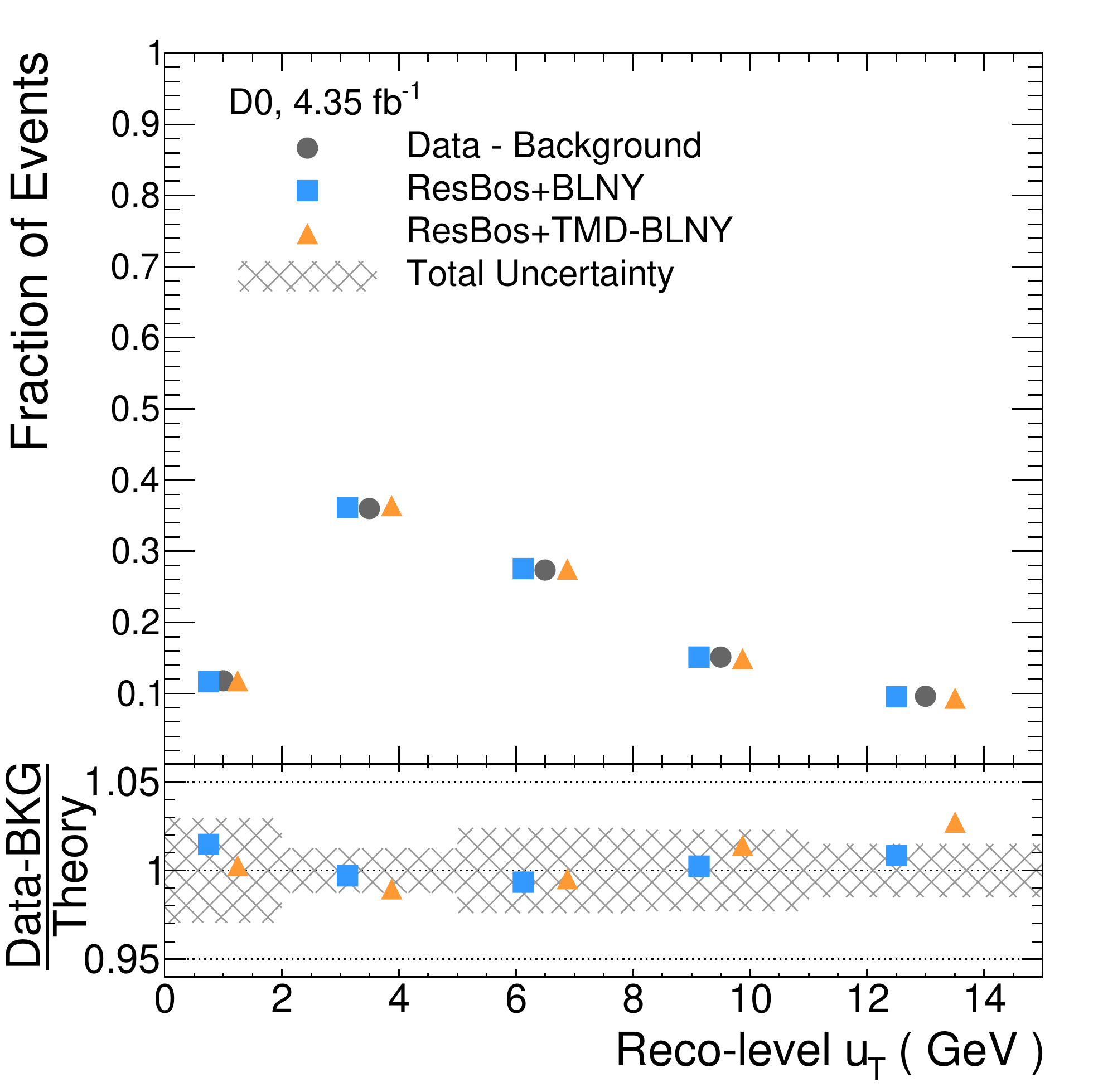}}
\hspace{0.1cm}
\includegraphics[width=0.51\columnwidth,height=6.8cm]{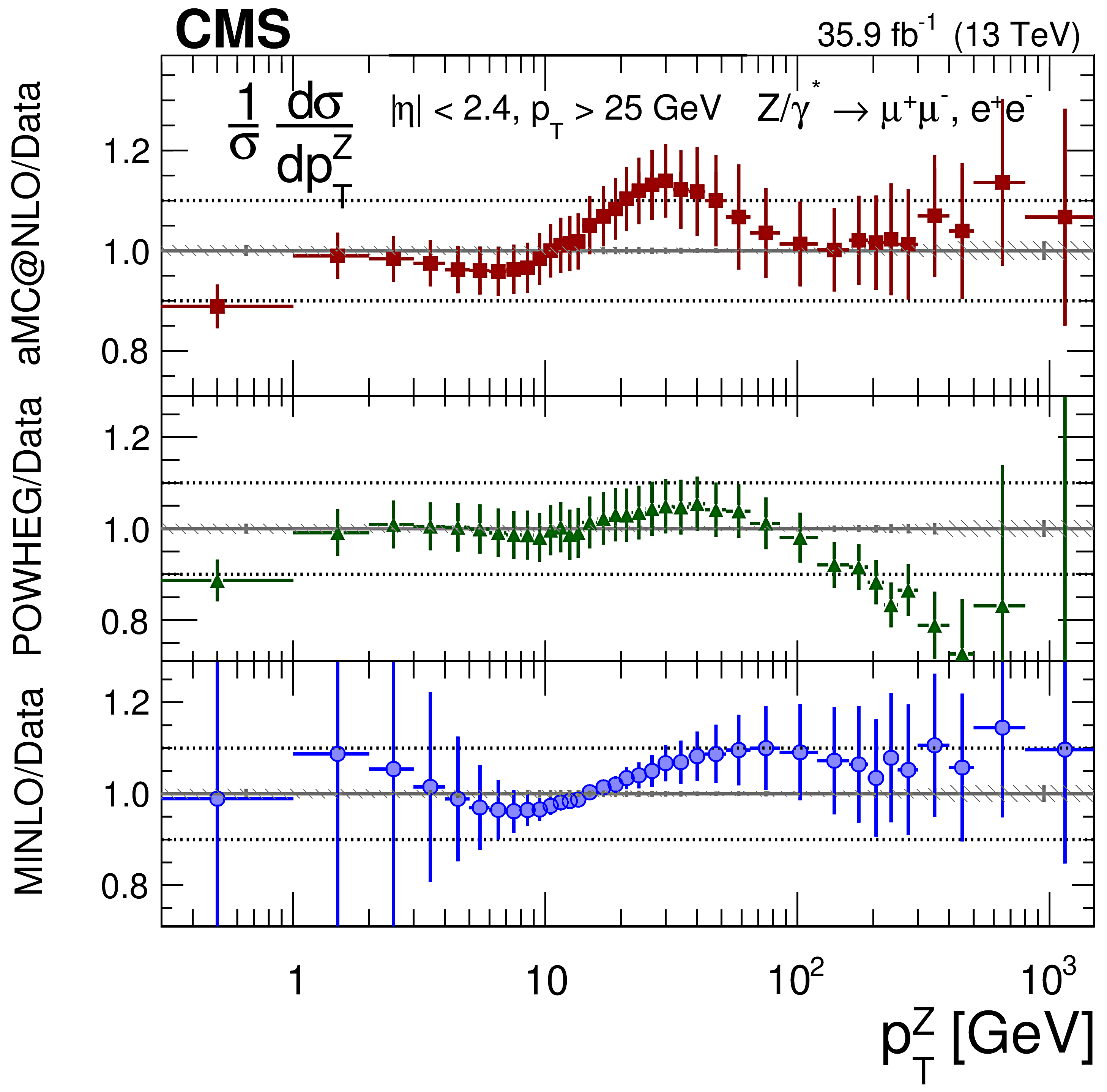}
\caption{Left: Hadronic recoil $u_{\rm T}$ in W-boson events measured by D0 compared to NLO+NNLL %\resbos\
predictions~\cite{Wang/D0,Abazov:2020moo}. Right: Ratio of theoretical to experimental Z-boson $\pt$ spectra in p-p %collisions 
at 13\,TeV~\cite{Gorbunov/CMS,Sirunyan:2019bzr}.}
\label{fig:NLL}
\end{figure}

\hspace{-0.5cm}The theoretical description of many pQCD differential cross sections requires the resummation of large soft and collinear logs appearing in processes sensitive to different energy scales. Typical examples are the spectra in the $\pt\to 0$ limit of massive particles at the LHC, such as the electroweak and Higgs bosons or heavy quarks, for which resummation of leading-log (LL) $\alphas\log(\pt/m)$ terms is needed.
Different measurements were presented confronted to N$^{\rm n}$LL resummed calculations:
\begin{itemize}
\item Low-$\pt$ W boson: The D0 experiment showed dedicated measurements of the hadronic recoil in W events (normalized to the accurately calibrated soft Z boson spectrum) compared to various models~\cite{Wang/D0,Abazov:2020moo}. The NLO+NNLL \resbos\ calculations are within a few \% of the data (Fig. \ref{fig:NLL}, left), whereas various \pythia8 tunes are excluded or disfavored. Such results provide important benchmarks for high-precision W mass extractions. %with different non-perturbative functions, BLNY and TMD-BLNY
\end{itemize}
\begin{itemize}
\item Low-$\pt$ Z boson: The full Z spectrum measured by CMS was confronted to various predictions confirming the sensitivity of different $\pt$ regions to fixed-order, resummed, and parton shower calculations (Fig. \ref{fig:NLL}, right)~\cite{Gorbunov/CMS,Sirunyan:2019bzr}. Resummed calculations based on TMD factorization~\cite{Taheri} are shown to reproduce well the softest part of the Z\,$+$\,jets spectra at the LHC as well as of the Drell--Yan data in p-p collisions at lower $\sqrts$.
\item Low-$\pt$ HQ: Spectra of charm and bottom mesons measured by ALICE~\cite{Vermunt/ALICE} and CMS~\cite{Mariani/CMS} are found consistent with, but systematically above, FONLL predictions at low $\pt$, suggesting the need to add missing NNLL contributions.
\end{itemize}

%%%%%%%%%%%%%%%%%%%%%%%%%%%%%%%%%%%%%%%%%%%%%%%%%%%%%%%%%%%%%%%%%%%%%%%%%%%%%%%%%%%%%%%%%%%%%%%%%%%%%%%%%%%%
\section{Parton shower and jet substructure}

Advanced studies of the energy-angle (sub)emissions within a jet have been performed in the last years exploiting modern jet substructure techniques~\cite{Larkoski:2017jix}. Intrajet parton emissions are characterized by their relative momentum $z$ and radius (angle) $R$ with respect to the jet, and visualized in the $(\ln(1/z),\ln(R/\Delta R))$ Lund plane that can be
%is a visual representation of the (hard collinear, hard wide angle, soft collinear, non-perturbative) phase space
constructed through repeated Cambridge/Aachen declustering of individual jets to follow the underlying parton splittings~\cite{Dreyer:2018nbf}. The soft-drop grooming is a popular algorithm that steps through the branching history removing branches that fail given $\pt,R$ requirements. The following experimental jet substructure results were shown at ICHEP'20:
\begin{itemize}
\item Inclusive jets: ATLAS presented different projections of the Lund plane variables ($\ln(1/z)$, $\ln(R)$, jet mass), and compared their various phase space regions to analytic (NLO+NLL, with and without non-pQCD corrections) and PS models~\cite{Roloff/ATLAS}. CMS jet mass studies in dijet events show that grooming reduces uncertainties by removing %contamination from 
soft particles and pileup, and that PY8 alone seems to reproduce better the data than \powheg+PY8 and \herwig++~\cite{Sunar/CMS}.
\end{itemize}
\begin{figure}[htbp!]
\centering
\includegraphics[width=0.49\columnwidth]{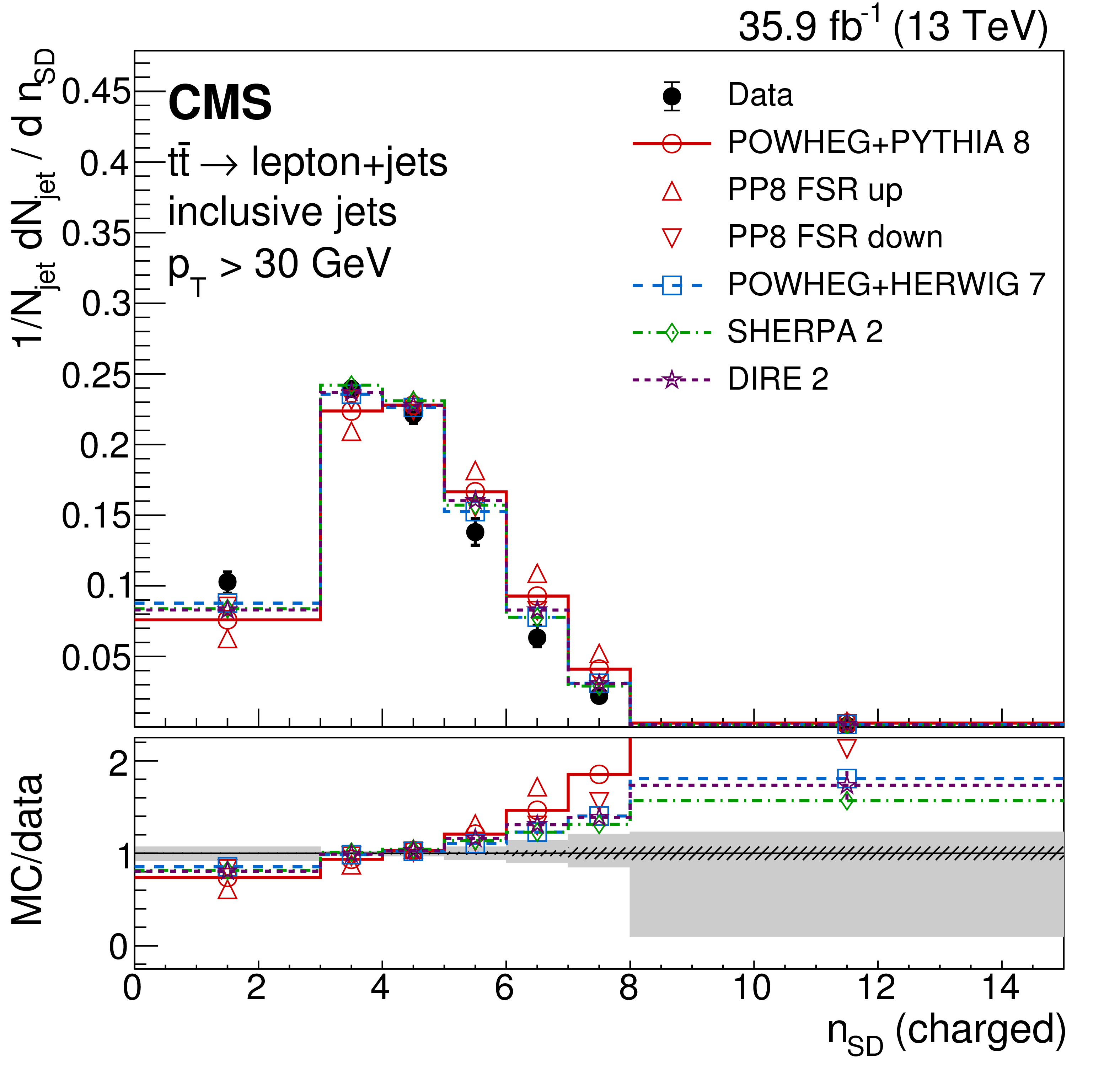}
\hspace{0.1cm}
\includegraphics[width=0.49\columnwidth]{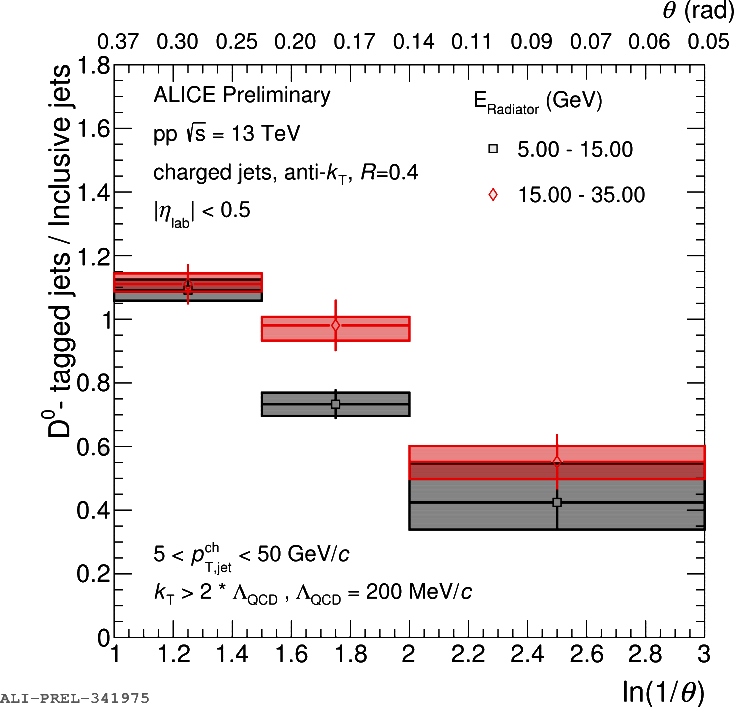}
\caption{Left: Distribution of the soft-drop multiplicity for inclusive jets measured in the CMS $\ttbar$ data compared to MC predictions~\cite{Sunar/CMS,Sirunyan:2018asm}.
Right: Ratio of angular distributions of splittings for $D^0$-tagged over inclusive jets measured by ALICE, showing depleted emissions for increasingly small angles~\cite{CunqueiroMulligan/ALICE}.}
\label{fig:subst}
\end{figure}
\begin{itemize}
\item Flavoured jets: Multiple jet substructure observables (generalized angularities, groomed momentum fraction, N-subjettiness ratios,...) measured by CMS for jets of different flavours produced in top pair events, indicate 10--50\% data-model differences (Fig.~\ref{fig:subst}, left)~\cite{Sunar/CMS,Sirunyan:2018asm}. These results call for improved theoretical developments of parton showering.
\item Charm jets: ALICE applied grooming techniques to inclusive and charm-tagged jets confirming the harder charm fragmentation compared to light-quarks and gluons, and the presence of suppressed radiation at small angles for $c$ quarks (Fig.\,\ref{fig:subst}, right)~\cite{CunqueiroMulligan/ALICE}. This~latter result is the first direct observation of the predicted pQCD heavy-quark ``dead cone'' effect~\cite{Dokshitzer:1991fd}.
\end{itemize}

%%%%%%%%%%%%%%%%%%%%%%%%%%%%%%%%%%%%%%%%%%%%%%%%%%%%%%%%%%%%%%%%%%%%%%%%%%%%%%%%%%%%%%%%%%%%%%%%%%%%%%%%%%%%
\section{Semihard and soft scatterings}

\begin{figure}[htbp!]
\centering
\includegraphics[width=0.43\columnwidth]{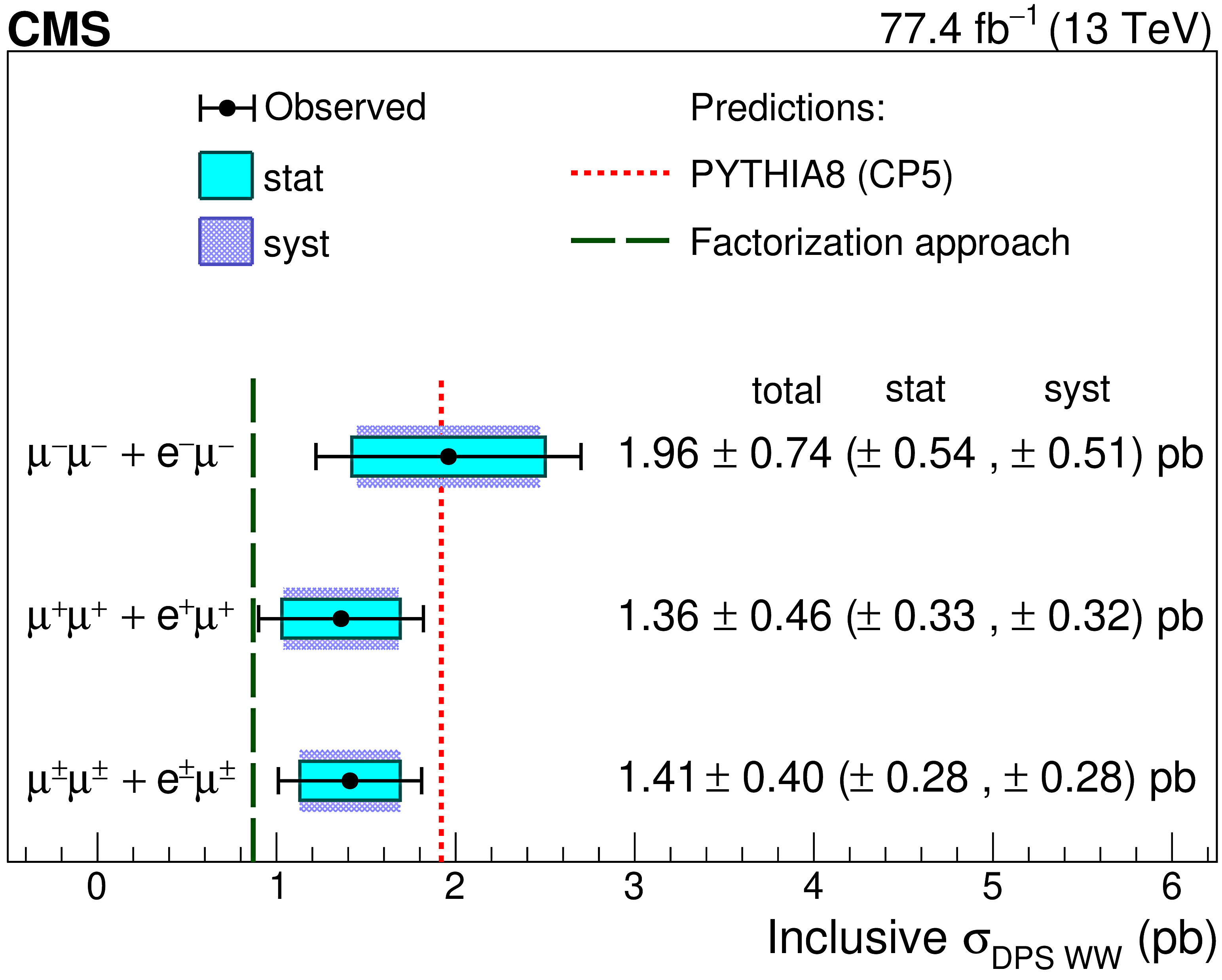}
\hspace{0.1cm}
\includegraphics[width=0.55\columnwidth]{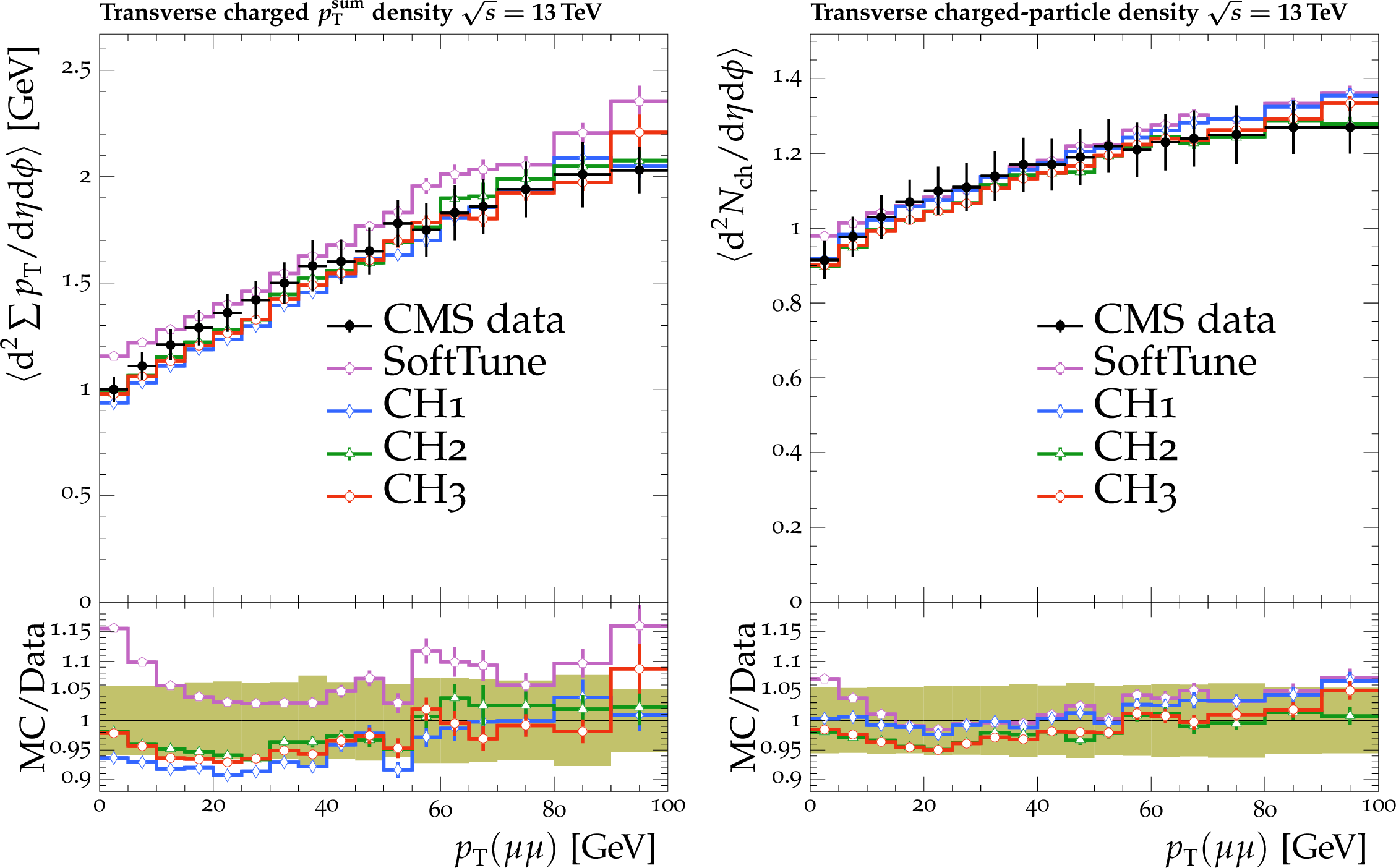}
\caption{Left: Cross section for DPS same-sign WW production compared to model expectations~\cite{Salvatico/CMS,Sirunyan:2019zox}. Right: UE activity measurements as a function Z-boson $\pt$ compared to various \herwig\ MC tunes~\cite{VanOnsem/CMS,Sirunyan:2020pqv}.}
\label{fig:semihard}
\end{figure}

Numerous LHC results were presented connected to studies of the strong interaction at semihard (few GeV) or soft (around $\lqcd\approx 0.2$\,GeV) scales. These included new measurements of double parton scatterings (DPS), multiparton interactions (MPI) of particular relevance to describe the underlying event (UE) in MC generators, and diffractive scatterings with hard or soft momentum exchanges:
\begin{itemize}
\item Double parton scatterings: CMS presented the first evidence ($3.9\sigma$) for the very rare same-sign WW process, historically considered a DPS ``smoking gun'' (Fig.~\ref{fig:semihard}, left), with associated effective DPS cross section of $\sigma_{\rm eff}\approx 12.7^{+5.0}_{-2.9}$~mb~\cite{Salvatico/CMS,Sirunyan:2019zox}. Double-$\Upsilon$ production was studied by CMS, showing that DPS contributions amount to 35\% of the inclusive yields~\cite{Amaral/CMS,Sirunyan:2020txn}, and the $\jpsi$+W results from ATLAS indicate $\sigma_{\rm eff}\approx 6$~mb~\cite{Abbot/ATLAS,Aaboud:2019wfr}. All these results provide novel information on the transverse parton profile and the parton correlations inside the proton. %probes different parton flavour transverse profile than double-QQ. Effective pp x-section tested: sigmaeff = 6--12 mb (lower value preferred)
\item Multiparton interactions: CMS showed that updated NLO+PY8 tunes based on minimum-bias data can also consistently reproduce the UE activity of multiple hard-scattering final states, and provided also the first set of \herwig\ MC tunes (Fig.~\ref{fig:semihard}, right)~\cite{VanOnsem/CMS,Sirunyan:2020pqv}. Similarly, the new ATLAS Z\,$+$\,jet UE data confirmed the need of retuning the MPI parameters of previous NLO+PY8, \sherpa, and \herwig\ simulations~\cite{Staszewski/ATLAS}.
\end{itemize}
\begin{itemize}
\item Diffractive and elastic scatterings: The first measurement of Mueller-Tang jet-gap-jet events presented by CMS shows a $f_{\rm gap}=0.6$--1.0\% rapidity-gap survival probability, with differential distributions not well reproduced by models~\cite{Baldenegro/CMS}. Cross sections of single-diffraction dijets with a forward proton tag have $f_{\rm gap}\approx7\%$ and are consistent with the \pythia 8 DG and \textsc{Pomwig} models, but overestimated by \pythia 8 4C and CUETP8M1 tunes~\cite{Suranyi/CMS}. ATLAS presented a precise measurement of the pomeron intercept and elastic slope in single-diffractive events tagged with forward protons~\cite{Staszewski/ATLAS}, while a comparative analysis of TOTEM and D0 elastic scattering data shows differences suggestive of the presence of odderon (colorless 3-gluon) exchanges at the LHC~\cite{Royon/CMS}. New measurements of $\rho$ photoproduction and central-exclusive dipion production were shown by the H1 and STAR experiments~\cite{Boltz/H1,Sikora/STAR,Truhlar/STAR}.
\end{itemize}

%%%%%%%%%%%%%%%%%%%%%%%%%%%%%%%%%%%%%%%%%%%%%%%%%%%%%%%%%%%%%%%%%%%%%%%%%%%%%%%%%%%%%%%%%%%%%%%%%%%%%%%%%%%%
\section{Parton hadronization}

\begin{figure}[htbp!]
\centering
\raisebox{10pt}{\includegraphics[width=0.59\columnwidth]{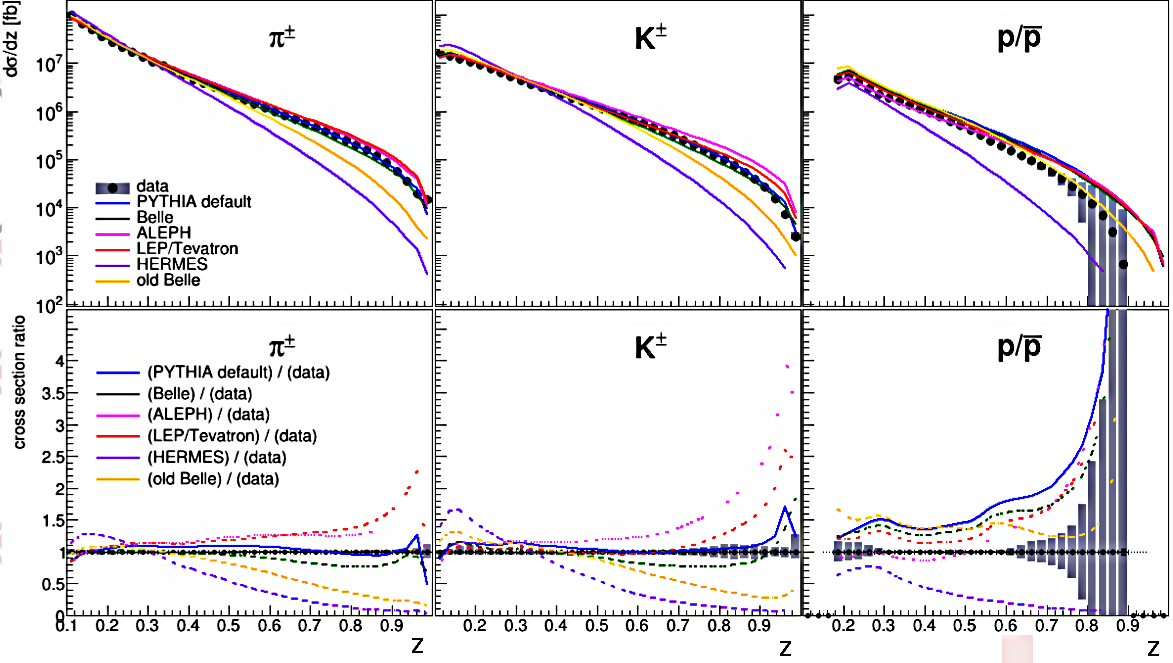}}
\hspace{0.1cm}
\includegraphics[width=0.39\columnwidth]{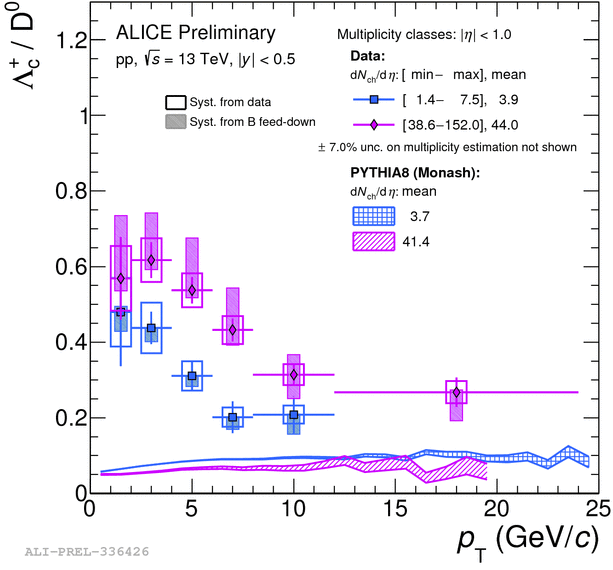}
\caption{Left: Fragmentation functions of $\pi^\pm$, K$^\pm$ and (anti)protons measured in $\epem$ collisions at $\sqrts = 10.58$\,GeV compared to model expectations~\cite{Seidl/BELLE,Seidl:2020mqc}. Right: Ratio of open charm baryon-to-meson ($\Lambda_c/$D$^0$) production vs.\ $\pt$ for two multiplicities in p-p at $\sqrts = 13$\,TeV compared to \pythia8 simulations~\cite{Zhu/ALICE}.}
\label{fig:FFs}
\end{figure}

New experimental studies of light-, heavy-quark, and gluon fragmentation functions (FFs) into mesons and baryons in $\epem$ and p-p collisions were presented. The data, in particular for baryons and for increasingly central p-p collisions with large particle multiplicities, strongly challenge the conventional Lund string hadronization in the ``QCD vacuum'', implemented \eg\ in \pythia8. The following hadronization studies were shown at ICHEP'20:
\begin{itemize}
\item Light-flavour hadrons: High-precision fragmentation functions of quarks into pions, kaons, and (anti)protons were presented by BELLE (Fig.~\ref{fig:FFs}, left)~\cite{Seidl/BELLE,Seidl:2020mqc}. The production of protons, in particular, remains challenging for MC models even in the QCD-free $\epem$ environment. ALICE presented high-$\pt$ $\pi^0$ and $\eta$ spectra over $\pt = 0.3$--200 GeV in p-p collisions at the LHC, showing that pQCD calculations using NLO FFs and \pythia8 have both difficulties reproducing the meson spectra over $\pt =2$--20 GeV, indicating inaccurate gluon-to-pion FFs and/or challenging their (process-independent) universality~\cite{Jonas/ALICE}.
\item Heavy-quark hadrons: The FFs of charm quarks were measured by ALICE, using leading D$^0$ mesons and $\Lambda_c$ baryons, finding a decent agreement with NLO \powheg+\pythia, except at low charm-jet $\pt$ where the FF is softer in data~\cite{Mazzilli/ALICE}. The production of charmed baryons ($\Lambda_c$, $\Sigma_c$, $\Xi_c$) was compared to the D$^0$ meson one as a function of hadron $\pt$ in ALICE~\cite{Zhu/ALICE}. Charm baryon yields are found to be enhanced by factors of 10--30 in some $\pt$ bins in p-p compared to $\epem$ collisions (Fig.~\ref{fig:FFs}, right). The fact that about 33\% (6\%) of charm quarks baryonize in p-p ($\epem$) requires strong colour reconnection and/or other collective final-state effects.
\end{itemize}
%\end{itemize}
%\begin{itemize}

%%%%%%%%%%%%%%%%%%%%%%%%%%%%%%%%%%%%%%%%%%%%%%%%%%%%%%%%%%%%%%%%%%%%%%%%%%%%%%%%%%%%%%%%%%%%%%%%%%%%%%%%%%%%
\section{Summary}

The precision needed to fully exploit the SM and BSM programs at the LHC requires an exquisite control of the physics of the strong interaction. A vast number of increasingly precise QCD observables are being experimentally studied and confronted to state-of-the-art theoretical predictions, thereby leading to further improved analytic calculations and parton shower MC generators. The experimental QCD studies presented at ICHEP'20, summarized here, covered multiple aspects such as new $\alphasmZ$ extractions, detailed data comparisons to N$^{\rm n}$LO and N$^{\rm n}$LL pQCD, improved PDFs, advanced jet substructure analyses, detailed parton radiation studies, and novel insights into semihard (double parton scatterings, underlying event,...) and non-perturbative (parton hadronization, colour reconnection,...) dynamics.

As a ``summary of a summary'', it is probably worth to highlight the following novel ICHEP'20 results: (i) tests of asymptotic freedom up to 4\,TeV, (ii) comparisons of multiple differential cross sections to the new theoretical standard defined by NNLO pQCD, (iii) first measurements of charm and bottom jets spectra at forward rapidity, (iv) improved constraints of the strange-quark PDF from W\,$+$\,jets,\,charm data, (v) precise measurements of the W-boson soft hadronic recoil (compared to NLO+NNLL calculations), (vi) the first observation of the heavy-quark dead-cone effect (exploiting jet substructure techniques), (vii) the first evidence of same-sign WW production from double parton scattering, and (viii) the significantly increased production of heavy-quark baryons in p-p compared to $\epem$ collisions that clearly challenges the universality of parton hadronization. All such interesting QCD measurements testify the strong vitality and sustained progress of the field.

%%%%%%%%%%%%%%%%%%%%%%%%%%%%%%%%%%%%%%%%%%%%%%%%%%%%%%%%%%%%%%%%%%%%%%%%%%%%%%%%%%%%%%%%%%%%%%%%%%%%%%%%%%%%

\end{document}